\documentclass[preprint]{elsarticle}

\usepackage{graphicx}
\usepackage[utf8]{inputenc}
\usepackage{hyperref}
\usepackage{amssymb}

\begin{document}
\title{Reverse Monte Carlo modelling: the two distinct routes of calculating the experimental structure factor}

\author{V. Sánchez-Gil$^{a,}$\corref{cor1}}
\ead{vsanchez@iqfr.csic.es}

\author{E.G. Noya$^a$}

\author{L. Temleitner$^b$}

\author{L. Pusztai$^b$}

\address[vsg,egn]{Instituto de Qu{\'\i}mica F{\'\i}sica Rocasolano, CSIC, Serrano 119, E-28006 Madrid, Spain}
\address[lt,lp]{Institute for Solid State Physics and Optics, Wigner Research Centre for Physics, P.O. Box 49, H-1525 Budapest, Hungary}

\cortext[cor1]{Corresponding author}

\date{\today}
\begin{abstract}
Two different Reverse Monte Carlo strategies, 'RMC++' and 'RMCPOW', have been compared for determining the microscopic structure of some liquid and amorphous solid systems on the basis of neutron diffraction measurements. The first, '$g(r)$ route', exploits the isotropic nature of liquids and calculates the total scattering structure factor, $S(Q)$, via a one-dimensional Fourier transform of the radial distribution function. The second, called 'crystallography' route, is based on the direct calculation of $S(Q)$ in the reciprocal space from the atomic positions in the simulation box. We describe these two methods and apply them to four disordered systems of increasing complexity. The two approaches yield structures in good agreement to the level of two- and three body correlations; consequently, it has been proven that the 'crystallography route' can also deal perfectly with disordered materials. This finding is important for future studies of liquids confined in porous media, where 
 handling Bragg and diffuse scattering simultaneously is unavoidable. 
\end{abstract}

\maketitle
\section{Introduction}
The Reverse Monte Carlo (RMC) method\cite{MoSi_1988_1_359} is a simple tool used for decades for elucidating the detailed atomic level structure of liquids and solids from scattering measurements. Over the past 25 years, RMC has been successfully applied to a wide variety of disordered materials that display structural disorder of varying extent: simple liquids\cite{howe93}, molten salts\cite{procroysoc-1990}, molecular liquids\cite{PRE_2005_72_031502,poth_12,temla2014}, water\cite{wikfeldt-2009} and aqueous  solutions\cite{hars-2012}, metallic\cite{kaban-2009} and covalent\cite{jovari-2007,jovari-2008} glasses. A separate class of applications has targeted 'disordered crystals' in which long range (crystalline) order and local (i.e., within the first coordination sphere) disorder are present simultaneously: examples may be crystals of silver and copper halides\cite{keen-1996,keen-2003} and of tetrahedral molecules\cite{poth-2013}. 

It was clear early on\cite{keen-1996} that dealing with genuine crystalline materials requires strategies different from those applicable for isotropic liquids/amorphous materials, due to the presence of long range periodic symmetries and the locally anisotropic nature of crystals. Just before the turn of the millennium, the (so far) ultimate solution was created: the RMCPOW software\cite{Mellergaard1999} is able to calculate Bragg- and diffuse scattering intensities directly from the particle coordinates, even for powder diffraction data obtained from laboratory X-ray sources and thermal neutron diffraction. For experimental data measured over very wide momentum transfer ranges, the RMCProfile strategy\cite{tucker-2001}, that involves the separation of the Bragg profile and Fourier-transform to real space, and a subsequent modelling of the total radial distribution function and the Bragg-profile, is also frequently used. The PDFGui software\cite{pdfgui}, performing PDF-based analysis of powder diffraction data, is a powerful tool for providing structural models based on the radial distribution function of crystalline materials. This is an alternative to the strictly unit-cell based investigation of crystalline structures; on the other hand, it is not capable of dealing with genuinely disordered structures. For isotropic disordered materials the original strategy of RMC\cite{MoSi_1988_1_359} may be used, i.e., from the atomic positions, first the radial distribution functions (RDF) are calculated, which later are Fourier transformed to the reciprocal space, so that primary experimental information, the total scattering structure factor (TSSF) may serve as 'target function' of RMC. Software that can realize this strategy may be RMC++\cite{gereben-2007}, RMC\_POT\cite{gereben-2012} or RMCProfile\cite{tucker-2001,dove-2002}. Details of the two strategies will be provided below; for now, it is important to state that a proper comparison between the two strategies is still missing.
  
The primary aim of this work is to test these strategies for several model systems. Since it is obvious that the simple 
route, via the calculation of the RDF, cannot be applicable for crystals, what needs to be tested is whether 
the more time consuming 'crystallographic' approach\cite{Mellergaard1999} can be used for isotropic disordered systems, 
such as liquids. Beyond the 'per se' interest, the timeliness of such a study lies in that a very important class of 
'mixed' systems, 'fluids in pores' would require a method that can handle both perfect crystals (like zeolites) and 
liquids (like water)\cite{Lang_1993_9_1846,ChemRev_2008_108_4125,sanchez-2014}. Note that the 'crystallographic' approach has already been proven to reproduce the atomic structure of simple adsorbed fluids (up to the level of three body correlations) in zeolites of varying pore sizes using the N-RMC method in which the number of particles is an additional adjustable parameter\cite{sanchez-2014}. In that work the target structure factor was obtained by simulation rather than from experiments and the study was restricted to simple fluids. 
Structural investigations of such complicated materials, that are of utmost significance in catalysis, oil industry, 
soil chemistry..., will not 
be possible until an established method of structural modelling can be proven to be applicable. Our aim now is to see whether the 'crystallographic' approach is also suitable for fitting experimental structure factors for more complex fluids.

Bearing in mind the above, the two approaches are tested on disordered one component systems of increasing complexity, from liquid argon to amorphous silicon. Liquid argon (l-Ar) is one of the simplest fluids in all respects: it can be easily described using radially symmetric pair potentials\cite{rahman-1964}. Liquid gallium (l-Ga), is a unique metallic element with possible short-lived covalent bonds that manifest in the slightly unusual shape of the main peak of the total scattering structure factor\cite{bellisent-funel}. Liquid selenium (l-Se) is one of the most unusual elemental liquids, because of the twofold coordination of the atoms and the resulting chain-like structure\cite{jovari-2001}. Finally, 
amorphous silicon (a-Si) can be regarded as a classic example of a disordered fourfold-coordinated covalent material that, in contrast to its well-known crystalline form, lacks the long-range order\cite{gereben-1994}. In the cases of l-Ga, l-Se and a-Si, experimental data\cite{bellisent-funel,l_se-data,kugler-1993} are from neutron diffraction measurements. In the case of argon, a computer-generated model of the liquid\cite{white_1999} has been employed, for two reasons: (1) this way, no systematic experimental errors had to be cared for, and (2) the early experimental data\cite{yarnell-1973} exhibited some residual systematic errors that made a thorough comparison of the methods somewhat cumbersome.

\section{The two approaches for calculating the measurable total scattering structure factors within RMC}
\label{method}
 Details of the RMC method can be found in various publications\cite{MoSi_1988_1_359,tucker-2001,dove-2002,mcgreevy-2001,evrard-2005,gereben-2007,gereben-2012} and therefore, here we will concentrate only on the parts relevant for calculating the structure factor from particle coordinates.

In short, the RMC algorithm produces sets of three-dimensional particle coordinates for which the calculated structure factor fits the input diffraction data within the estimated experimental errors. The goodness-of-fit is quantified using a $\chi^2$-value:

\begin{equation}
\chi^2= \sum_{i=1}^{N_Q} \frac{(S_{calc} (Q_i) - S_{exp} (Q_i))^2}{\sigma^2 (Q_i)}  ,
\end{equation}
\\
where $Q$ is the modulus of the scattering variable, the sum runs over all experimental points, $N_Q$; $S_{exp}$ and $S_{calc}$ are the experimental and simulated structure factors, respectively, and $\sigma$ is the 'estimated' standard deviation for the experimental point $i$.

To minimize $\chi^{2}$, random movements are attempted for all atoms in the simulation box. If the new non-overlapping position reduces differences between experimental and calculated structure factors, the move will be accepted. Otherwise, the move is accepted according to an acceptance probability, $P^{acc}$, given by	
	
\begin{equation}
\label{pacc}
P^{acc} = \min \left( 1, \exp\left(-\frac{\chi_{new}^2-\chi_{old}^2}{2}\right) \right) ,
\end{equation}	
\\	
	
where $\chi_{old}^2$ and $\chi_{new}^2$ correspond to the original and proposed atomic coordinates, respectively.
	
Finally, an exclusion core around each particle is defined, $r_{cutoff}$, to reflect its effective size. If the proposed position overlaps with any other particle in the simulation box then the move will be automatically rejected. Further constraints can be applied, for example, on the coordination number and/or nearest neighbor distances and angles\cite{mcgreevy-2001,evrard-2005,gereben-2007,tucker-2001,dove-2002}.

The different approaches that we present here, are based on two different ways of calculating the total scattering structure factor , $S_{calc}$, from the particle coordinates.

\subsection{Method I: the 'g(r) route' (RMC++)}
~

This approach is based on the one-dimensional Fourier transformation of the radial distribution function (RDF). For one component systems, the RDF can simply be calculated from the atomic positions as 

\begin{equation}
\label{eq_RDF}
g(r) = \frac{n(r)}{\Delta V \rho},
\end{equation}

where $n(r)$ is the number of atoms at a distance between $r$ and $r+\Delta r$ from a central atom, $\Delta V$ is the volume of a spherical shell between  $r$ and $r+\Delta r$ and $\rho$ is the number density of the system.

Liquids and amorphous materials can be considered isotropic beyond nearest-neighbor distances so that for switching between the real and reciprocal space, a one-dimensional Fourier transform is widely used. Radial distribution functions can be Fourier transformed and weighted for the actual experiment thus providing the total scattering structure factor, $S(Q)$. For neutron scattering measurements and one component systems, the appropriate Fourier transform is given by

\begin{equation}
\label{eq_SQ_byFT}
S(Q) = 1 + \frac{4\pi\rho\langle b\rangle^2}{Q}\int_0^\infty{r [g(r)-1] sin(Q r) dr},
\end{equation}

where $\rho$ denotes the number density of the sample, $\langle b\rangle$ is the neutron scattering length of the atom type in question, $Q$ are the moduli of the reciprocal lattice vectors and the integral runs over atomic distances $r$. In practice, a discrete integration using the so called rectangular method\cite{evrard-2005} is performed with a summation whose upper limit is restricted by the half-length of the simulation box. This method is implemented in, for instance, the RMC++\cite{evrard-2005,gereben-2007}, RMC\_POT\cite{gereben-2012} and RMCProfile\cite{tucker-2001,dove-2002} software packages.

\subsection{Method II: the 'crystallography route' (RMCPOW)}
~

In contrast to Method I, the 'crystallography route', implemented by the software RMCPOW\cite{Mellergaard1999}, is based on the super-cell approximation, repeating the 'unit cell' (i.e., in our case, the simulation cell) in each direction. The total scattering structure factor, $S(Q)$, is calculated using a three-dimensional Fourier transformation to the reciprocal space from atomic coordinates. In this way, RMCPOW can deal with ordered and disordered systems because diffuse (local disorder) and Bragg scattering (crystalline, long range order) are both considered. Diffuse intensities, that are assumed to vary smoothly, are locally averaged whereas for Bragg intensities the same summation is performed without averaging (see Ref.\cite{Mellergaard1999} for details).
	
In the case of neutron diffraction, the orientationally averaged structure factor\cite{Mellergaard1999} is
\begin{equation}
\label{eq_sq}
S (Q) = \frac{ 2 \pi^2}{N V<b>^2} \sum_{{\mathbf Q'}} | F ({\mathbf Q'}) |^2  
\delta (  Q- Q')/Q'^2 .
\end{equation}

Where  $N$ and $V$ are, respectively, the number of atoms and the volume of system, ${\mathbf Q'}$ are the allowed vectors in the reciprocal cell, and $\langle b\rangle$ is the average of the coherent scattering lengths. The $1/{Q'}^2$ factor stems from the angular integration over all the possible ${\bf Q}'$ orientations\cite{Mellergaard1999}. $F({\mathbf Q})$ contains the correlations between scattering nuclei and is given by
		
\begin{equation}
	F({\mathbf Q}) = \sum_{j=1}^{N} b_j \exp(i{\mathbf Q}{\mathbf R_j}) ,
\end{equation}

where ${\mathbf R}_j$ denotes the position of atom $j$ in the unit cell. 

It is important to point out that no Fourier transformation is involved in this scheme and therefore, the usual numerical problems (truncation, aliasing) in conjunction with that do not occur. Another thing to notice is that if Eq.\ref{eq_sq} was calculated for an isotropic system without periodical long range ordering then the vectors could be substituted by their magnitudes and the summation could be replaced by integration; that is, eventually, Eq.\ref{eq_SQ_byFT} would be reproduced.

\section{Calculations performed}

\begin{figure}[!h]
\begin{center}
\includegraphics[width=120mm,angle=0,clip=true]{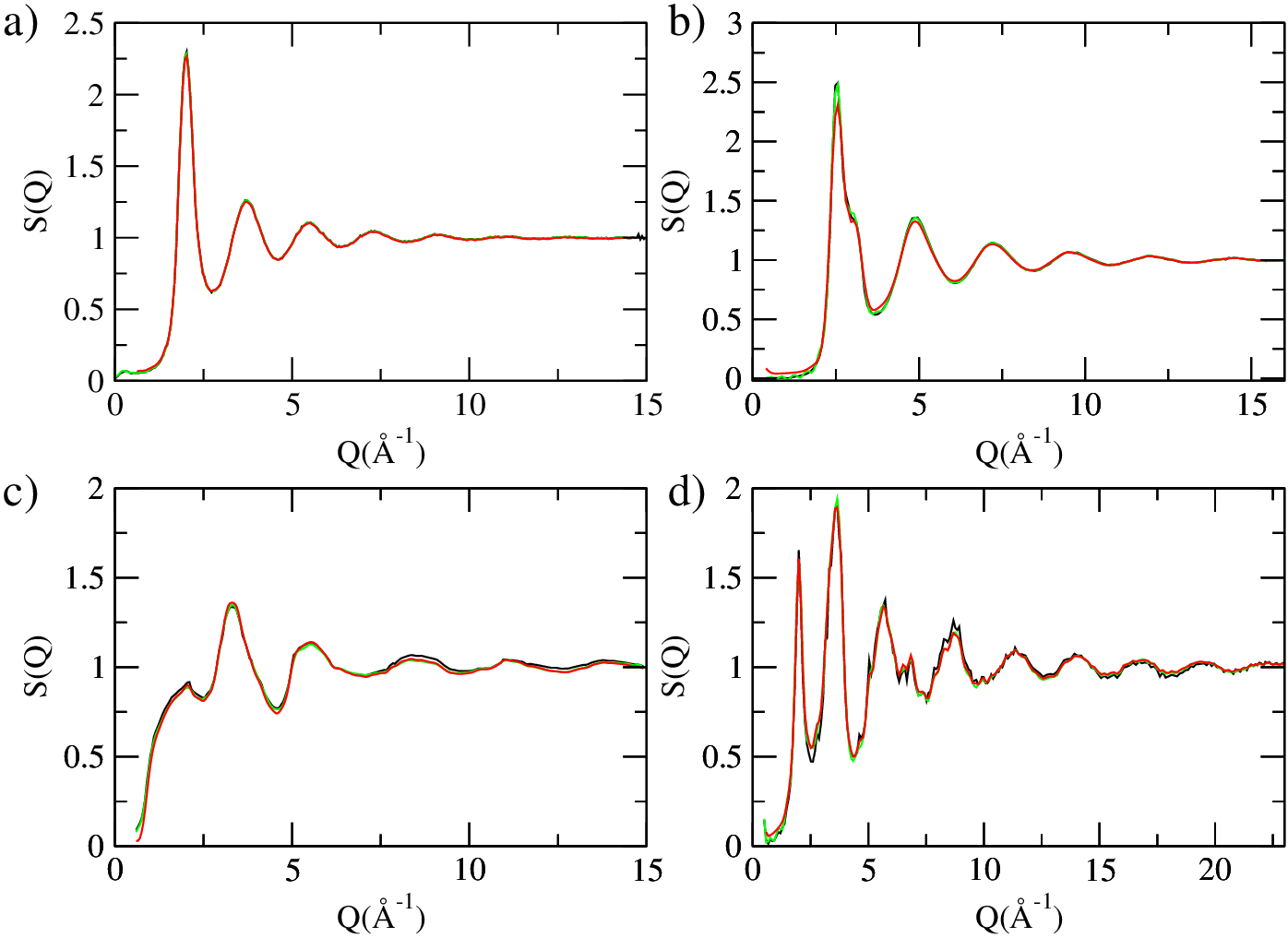}
\caption{\label{figure_structure_factor} Comparison of the experimental
structure factors provided by experiments (black line), RMC++ (green line) and RMCPOW (red line). From top to bottom and from left to right a) liquid argon (modelled data, from Monte Carlo simulation, see text for details) b) liquid gallium (experiment from Ref.\cite{bellisent-funel}) c) liquid selenium (experiment from Ref.\cite{l_se-data}) d) amorphous silicon (experiment from Ref.\cite{kugler-1993}).}
\end{center}
\end{figure}

As mentioned before, approaches I and II are tested here on disordered one component systems of increasing complexity, from liquid argon to amorphous silicon. The differences in terms of structural order can be clearly seen in Fig.\ref{figure_structure_factor}, where the experimental and simulated structure factors for the four systems are shown. Note that as the complexity of the test systems increases, new features of the 'diffuse' scattering appear but not any Bragg peak and therefore method I (the '$g(r)$ route') can also be used. In all cases, simulated and experimental data are from neutron diffraction 'measurements'. 

For l-Ar, the simplest case, modelled data from canonical Monte Carlo simulations at 85K have been included. In this way, the target structure is accurately known and we have access to the real RDF and ADF to compare with. In the canonical Monte Carlo simulation argon atoms are modelled using Lennard-Jones interactions and the parameters of the LJ potential were taken from the literature\cite{white_1999}.

All simulations have been performed using the RMC++\cite{gereben-2007} and RMCPOW\cite{Mellergaard1999} free software packages with cubic simulation boxes of side length 32\AA{} for l-Ga and a-Si and 50\AA{} for l-Ar and l-Se. In table \ref{tbl_parameters} the experimental density and the effective size of the particles for each test case are shown. For simplicity, as the aim of this work is to compare the two approaches, we have chosen a uniform value for the 'experimental' standard deviation of all $Q$-values, $\sigma$=0.001. As it can be seen in Fig.\ref{figure_structure_factor}, where target and simulated structure factors are hardly distinguishable, this value produces good quality fits.

\begin{table}[!h]
\centering
\caption{\label{tbl_parameters} Experimental density and $r_{cutoff}$ used
for the model systems. }
\begin{tabular}{lcccc}
\\
\hline\hline
	& & $\rho$ (atoms/\AA$^3$)  & & $r_{cutoff}$  (\AA)  \\
\hline
l-Ar   & &   0.02125    & &     2.7  \\      
l-Ga   & &   0.05197    & &     2.2  \\      
l-Se   & &   0.0298     & &     2.0  \\      
a-Si   & &   0.04846    & &     2.2  \\
 \hline\hline
\end{tabular} \\
\end{table}

A better comparison of configurations provided by the different RMC strategies can be made by computing the radial distribution functions (RDF) and some simple three body correlation functions. Such comparison is able to reveal subtle variations of the structure that result from the different ways of calculating the total scattering structure factor. As defined above, the RDF can be determined from Eq.\ref{eq_RDF}. For three body correlations we calculate the bond angle distribution ('angular distribution function', ADF) that can be defined as the integral of the three body correlation function $g^{(3)}(r_1,r_2,\cos\theta)$ over the first coordination shell:
\begin{equation} 
\label{ftheta}
f (\theta) = 16 \pi^2 \int_0^{r_c} \int_0^{r_c} r_{13}^2 dr_{13} r_{23}^2 dr_{23} g(r_{12}) g^{(3)}(r_{13},r_{23},\cos\theta) ,
\end{equation} 

where we chose $r_c$ as the position of the first minimum of the radial distribution function in each test system. This function gives the distribution of angles between pairs of nearest neighbors with respect to a central atom. The neutron scattering lengths have been taken from Ref.\cite{scat_lengths}.

\section{Results and discussion}
\label{results}

\subsection*{•Liquid argon}

\begin{figure}[!h]
\begin{center}
\includegraphics[width=120mm,angle=0,clip=true]{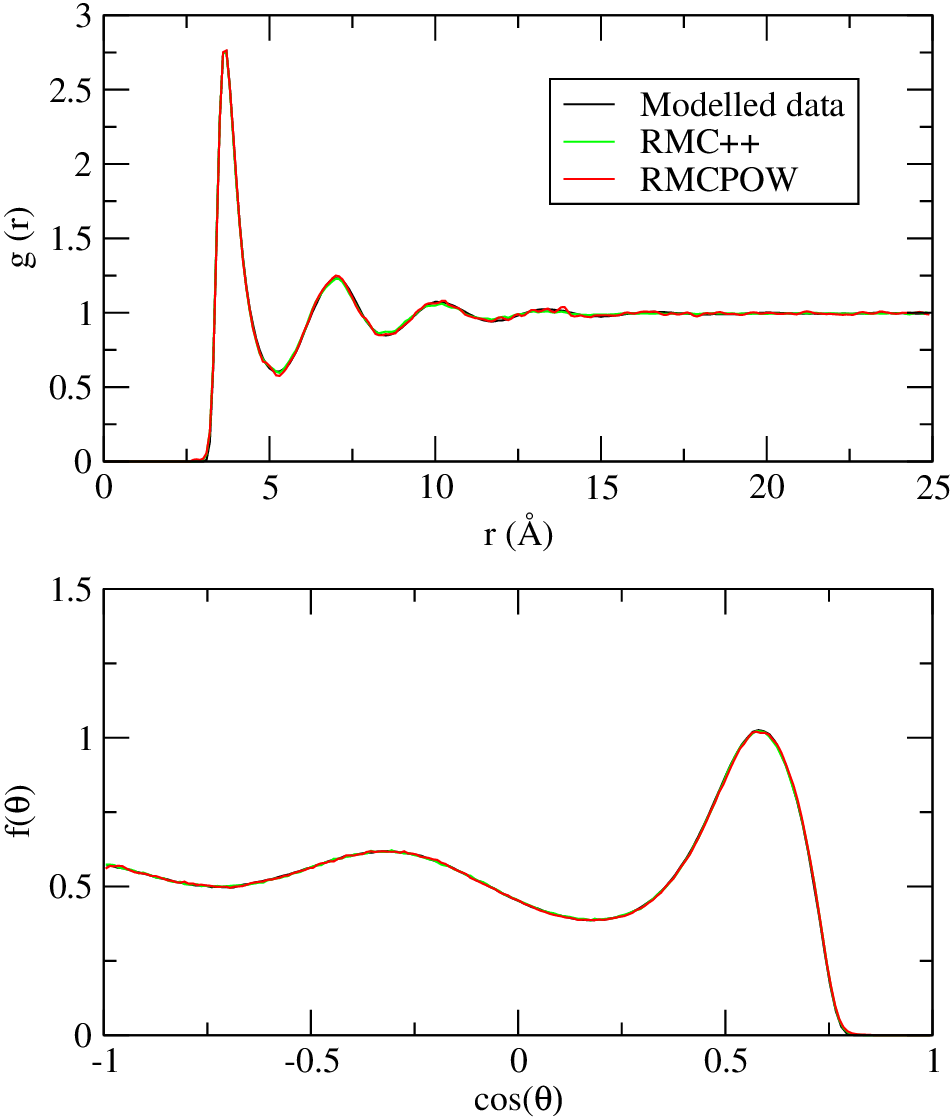}
\caption{\label{modelled_Ar} Comparison of the simulated radial distribution functions (top) and bond angle distribution functions (bottom) provided by canonical Monte Carlo simulations (black line), RMC++ (green line) and RMCPOW (red line) for liquid argon.}
\end{center}
\end{figure}

We start by presenting results for argon using the modelled data from the canonical Monte Carlo simulation. As it can be seen in Fig.\ref{modelled_Ar}, the agreement is almost perfect for RDF and ADF. Therefore it is clearly shown that for 'perfect' experimental data, and for a system with purely two body interactions, both RMC approaches reproduce the target structure to the level of two- and three-body correlations.

\subsection*{•Liquid gallium}

\begin{figure}[h!]
\begin{center}
\includegraphics[width=120mm,angle=0,clip=true]{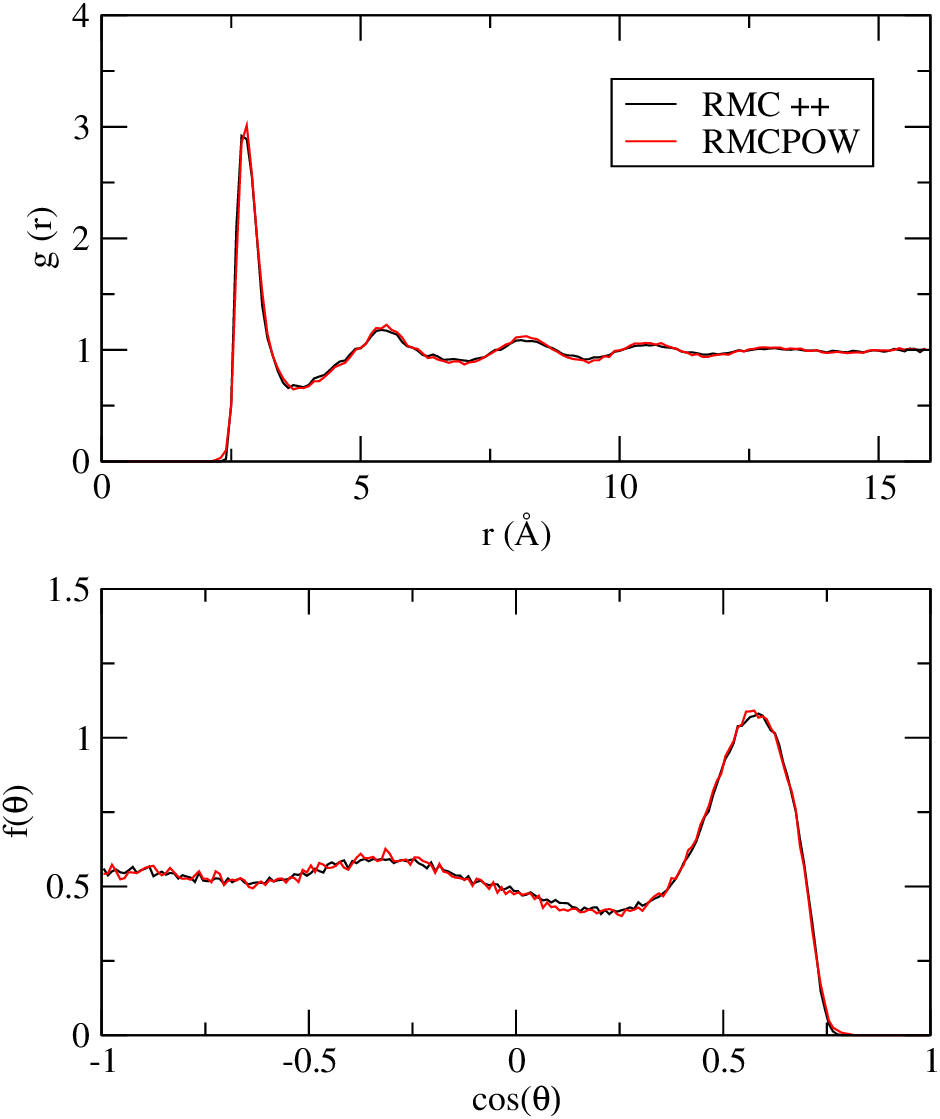}
\caption{\label{Gallium}  Comparison of the radial distribution function (top) and the bond angle distribution function (bottom) for liquid gallium.}
\end{center}
\end{figure}

In this case the agreement between the RMC++ and the RMCPOW radial distribution functions and bond angle distributions (see Fig.\ref{Gallium}) is almost perfect. Behind the good quality of the match of RDF-s and ADF-s one finds the considerably higher experimental density of l-Ga in comparison with l-Ar (see table \ref{tbl_parameters}) and the wider $Q$ range (up to 16\AA$^{-1}$) and better quality of the neutron scattering measurement. As a consequence of the higher density for gallium, the two approaches yield smooth RDF and ADF simply using a simulation box of side 32\AA.

\subsection*{•Liquid selenium}

\begin{figure}[!h]
\begin{center}
\includegraphics[width=120mm,angle=0,clip=true]{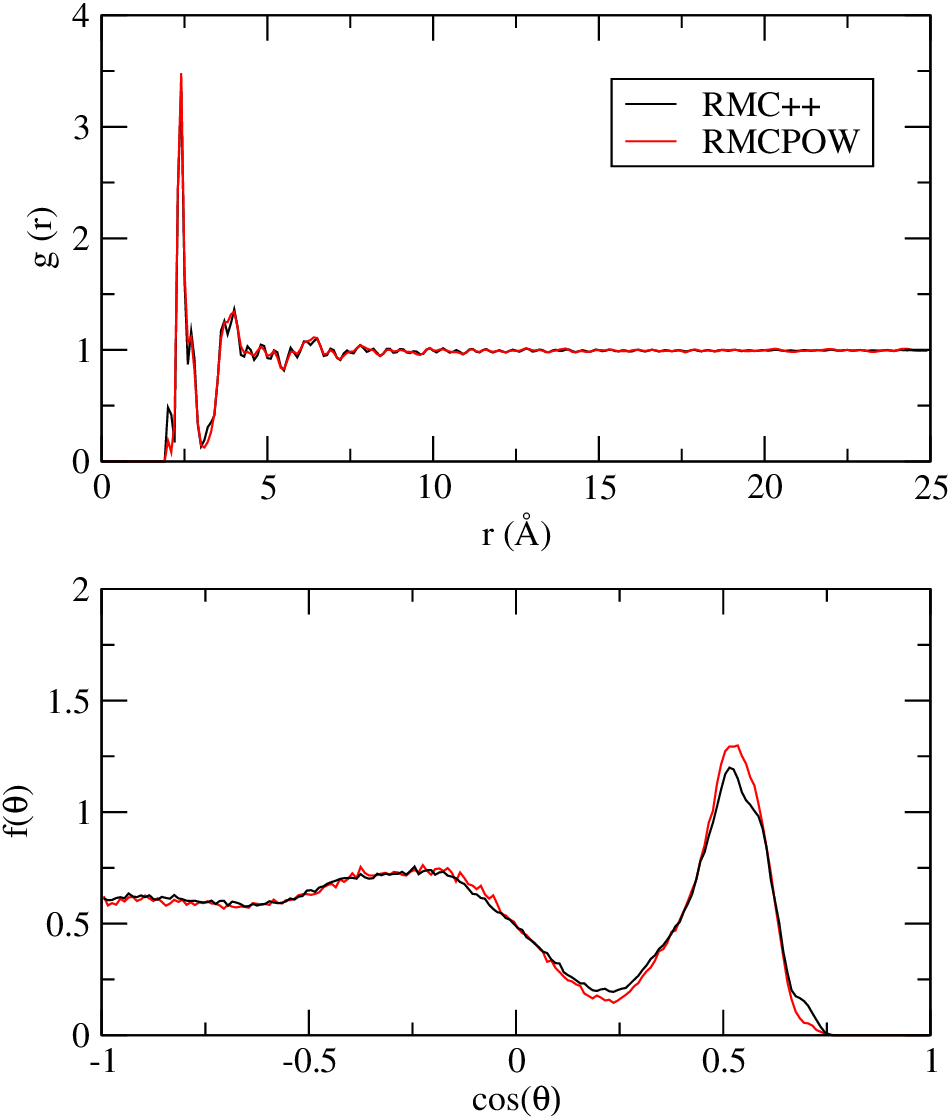}
\caption{\label{Selenium}  Comparison of the radial distribution function (top) and the bond angle distribution function (bottom) for liquid selenium.}
\end{center}
\end{figure}

For l-Se, when comparing the radial and bond angle distribution functions in Figure \ref{Selenium}, the overall good agreement is apparent, although the look of these functions is not as nice as it was for liquid gallium. A first glance at the RDF shows that the fluid is rather structured at short distance, because of the two covalent bonds of the atoms; this feature, however, does not seem to impose longer range ordering. The short period oscillations of the $g(r)$, again, are probably due to some residual systematic errors of the experimental $S(Q)$\cite{l_se-data}.

For systems like liquid selenium, in which the short range $g(r)$ displays significant features, fine long range details of $S(Q)$ cannot be neglected. Also, the size of the simulation box affects to the accuracy of both approaches. By increasing the simulation box up to 50\AA{}more particle distances are included in the RDF calculation for method I and more reciprocal vectors are included in the evaluation of $S(Q)$ for method II. This, in turn, implies both the use of a large system size that allows a finer sampling of $r$-space (method I) or $Q$-space (method II) and the inclusion of a rather long  $Q$-range in the fitting procedure. 

The bond angle distributions are very similar, exhibiting maxima at the same angle values. The small difference for the maximum at 60 degree angle is not unexpected for a liquid with a relatively complex structure (twofold coordination and chain like structure). Only imposing some additional constrains, a reasonable good prediction for the three body correlations from a RMC simulation can be obtained for systems with this level of order.

\subsection*{•Amorphous silicon}

\begin{figure}[!h]
\begin{center}
\includegraphics[width=120mm,angle=0,clip=true]{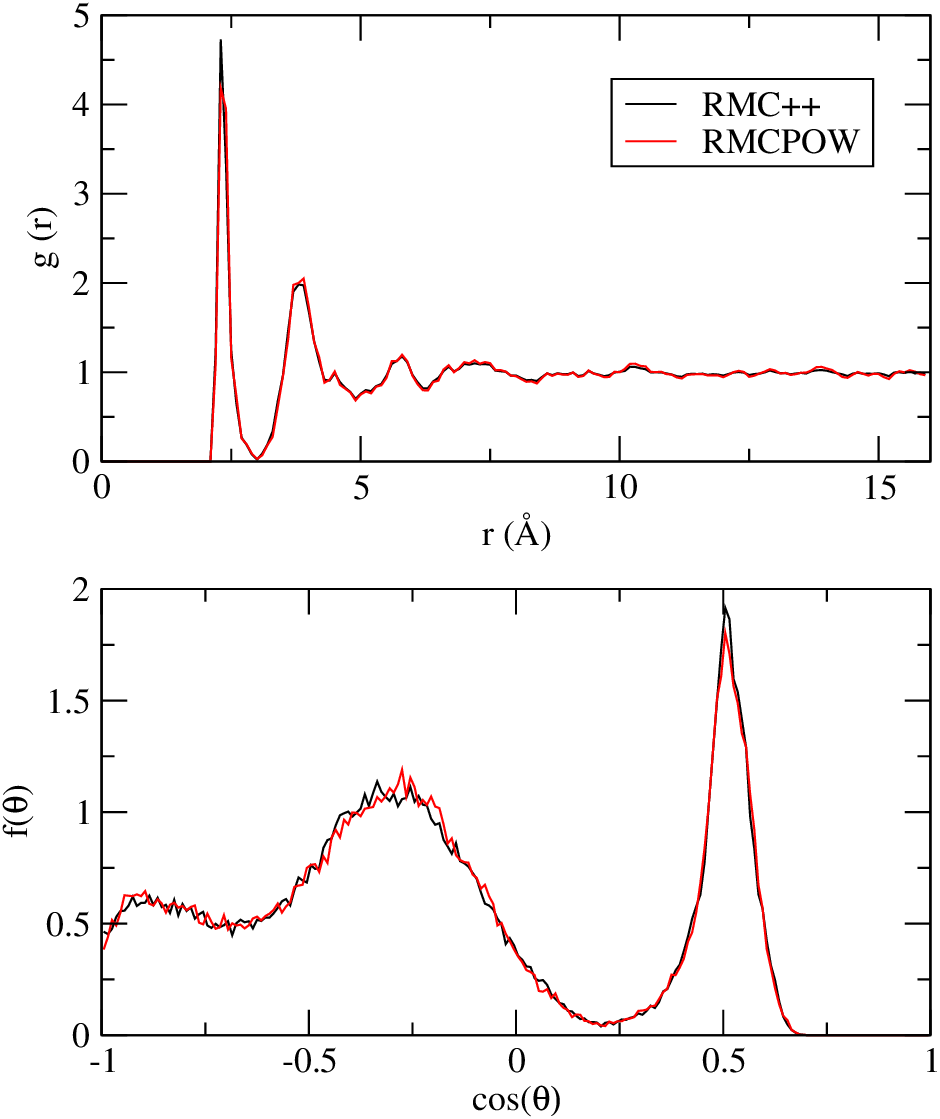}
\caption{\label{Silicon}  Comparison of the radial distribution function (top) and the bond angle distribution function (bottom) for amorphous silicon.}
\end{center}
\end{figure}

The RMC++ and RMCPOW radial distribution functions (see Fig.\ref{Silicon}) agree very well; that is, the highest level of ordering among our test systems has not posed particular difficulties to either approaches. Interestingly, the main maximum of the RDF resulting from the '$g(r)$ route' is slightly sharper than its counterpart. Since the total scattering structure factors belonging to RMC++ and RMCPOW run together, it is not possible to assess which RDF is the 'real' one: one must accept that both (only very slightly different) $g(r)$-s are possible solutions. It would also be rather hard to consider differences in terms the ADF-s significant: it might just be noticed that the unphysical maximum at the 60 degree angles is slightly stronger for the RMC++ solutions, whereas the 'real' maximum around the tetrahedral angle is very similar for the two approaches. 

\section{Summary and Conclusions}
\label{conclusion}

Two routes to the structure factor calculation implemented in the RMC++\cite{evrard-2005, gereben-2007} and RMCPOW\cite{Mellergaard1999} algorithms have been considered. They have been successfully tested and compared on four model systems with different level of (dis)order. The agreement is almost perfect for simple (l-Ar) and non-covalent non-simple (l-Ga) liquids even for more ordered systems (l-Se and a-Si), only minor differences appear in terms of the RDF-s and ADF-s. 

In terms of their relative efficiency, we found that the '$g(r)$ route' shows a considerably faster convergence, but it is exposed to Fourier truncation errors. Furthermore, this approach cannot be extended to deal with materials with Bragg scattering. In contrast, the 'crystallography' route can be used, at the expense of computational time, for a wide range of systems, from simple liquids to perfect crystals.

For two-phase systems of our future concern, porous crystalline materials with partially filled pores, 
i.e. in which crystalline and liquid/disordered phases are simultaneously present in the sample, the 'crystallography route' is the only approach. Out of the presently available software, the RMCPOW algorithm seems to be the most general choice; RMCProfile\cite{tucker-2001} may also be applicable of experimental data over extremely wide $Q$-range are available. This conclusion is also supported by a previous RMC study that showed that the 'crystallography route' provided an appropriate description of simple adsorbed fluids in zeolites, although in that case simulated target structure factors were used\cite{sanchez-2014}.

\section*{Acknowledgments}

V.S.G and E.G.N. gratefully acknowledge the support from the Spanish MINECO (Ministry of Innovation and Economy) Grant No. FIS2010-15502 and FIS2013-47350-C5-4-R. V.S.G. also acknowledges the CSIC for support of his work by means of a JAE program Ph.D. fellowship. L.T. an L.P. are grateful for the financial support provided by Grant No. 083529 of the Hungarian National Basic Research Fund (OTKA).


\bibliographystyle{elsarticle-num}

\begin{thebibliography}{10}
\expandafter\ifx\csname url\endcsname\relax
  \def\url#1{\texttt{#1}}\fi
\expandafter\ifx\csname urlprefix\endcsname\relax\def\urlprefix{URL }\fi
\expandafter\ifx\csname href\endcsname\relax
  \def\href#1#2{#2} \def\path#1{#1}\fi

\bibitem{MoSi_1988_1_359}
R.~L. McGreevy, L.~Pusztai, Reverse monte carlo simulation: A new technique for
  the determination of disordered structures, Mol. Simul. 1 (1988) 359.

\bibitem{howe93}
M.~A. Howe, R.~McGreevy, L.~Pusztai, I.~Borzsák, Determination of 3 body
  correlations in simple liquids by rmc modelling of diffraction data, Phys.
  Chem. Liq. 25 (1993) 205--241.

\bibitem{procroysoc-1990}
R.~McGreevy, L.~Pusztai, The structure of molten salts, Proc. Roy. Soc. London
  A 430 (1990) 241--260.

\bibitem{PRE_2005_72_031502}
N.~Veglio, F.~J. Bermejo, L.~C. Pardo, J.~L. Tamarit, G.~J. Cuello, Direct
  experimental assessment of the strength of orientational correlations in
  polar liquids, Phys. Rev. B 72 (2005) 031502.

\bibitem{poth_12}
S.~Pothoczki, A.~Ottochian, M.~Rovira-Esteva, L.~C. Pardo, J.~L. Tamarit, G.~J.
  Cuello, {Role of steric and electrostatic effects in the short-range order of
  quasitetrahedral molecular liquids}, Physical Review B 85 (2012) 014202.

\bibitem{temla2014}
L.~Temleitner, Structure determination of liquid carbon tetrabromide via a
  combination of x-ray and neutron diffraction data and reverse monte carlo
  modeling, Journal of Molecular Liquids 197 (2014) 204--210.

\bibitem{wikfeldt-2009}
K.~T. Wikfeldt, M.~Leetmaa, M.~P. Ljungberg, A.~Nilsson, L.~G.~M. Pettersson,
  On the range of water structure models compatible with x-ray and neutron
  diffraction data, J. Phys. Chem. B 113 (2009) 6246--6255.

\bibitem{hars-2012}
I.~Harsányi, L.~Pusztai, Hydration structure in concentrated aqueous lithium
  chloride solutions: A reverse monte carlo based combination of molecular
  dynamics simulations and diffraction data, J. Chem. Phys. 137 (2012) 204503.

\bibitem{kaban-2009}
I.~Kaban, P.~Jóvári, M.~Stoica, J.~Eckert, W.~Hoyer, B.~Beuneu, Topological
  and chemical ordering in {C}o$_{43}${F}e$_{20}${T}a$_{5.5}${B}$_{31.5}$
  metallic glass, Phys. Rev. B 79 (2009) 212201.

\bibitem{jovari-2007}
P.~Jóvári, I.~Kaban, J.~Steiner, B.~Beuneu, A.~Schöps, A.~Webb, 'wrong
  bonds' in sputtered amorphous {G}e$_2${S}b$_2${T}e$_5$, J. Phys. Condens.
  Matter 19 (2007) 335212.

\bibitem{jovari-2008}
P.~Jóvári, I.~Kaban, J.~Steiner, B.~Beuneu, A.~Schöps, M.~A. Webb, Local
  order in amorphous {G}e$_2${S}b$_2${T}e$_5$ and {G}e{S}b{T}e$_4$, Phys. Rev.
  B 77 (2008) 035202.

\bibitem{keen-1996}
D.~A. Keen, S.~Hull, W.~Hayes, N.~J.~G. Gardner, Structural evidence for a
  fast-ion transition in the high-pressure rocksalt phase of silver iodide,
  Phys. Rev. Lett. 77 (1996) 4914--17.

\bibitem{keen-2003}
D.~A. Keen, S.~Hull, A.~C. Barnes, P.~Berastegui, W.~A. Crichton, P.~A. Madden,
  M.~G. Tucker, M.~Wilson, The nature of the superionic transition in ag+ and
  cu+ halides, Phys. Rev. B 68 (2003) 014117.

\bibitem{poth-2013}
S.~Pothoczki, L.~Temleitner, L.~C. Pardo, G.~J. Cuello, M.~Rovira-Esteva, J.~L.
  Tamarit, {Comparison of the atomic level structure of the plastic crystalline
  and liquid phases of $CBr_2Cl_2$ : neutron diffraction and reverse Monte
  Carlo modelling}, Journal of Physics: Condensed Matter 25 (2013) 454216.

\bibitem{Mellergaard1999}
A.~Mellergård, R.~L. McGreevy, Reverse monte carlo modelling of neutron powder
  diffraction data, Acta Crys. A 55 (1999) 783 -- 789.

\bibitem{tucker-2001}
M.~G. Tucker, M.~T. Dove, D.~Keen, Application of the reverse monte carlo
  method to crystalline materials, J. Appl. Crystallogr. 34 (2001) 630--8.

\bibitem{pdfgui}
C.~L. Farrow, P.~Juhás, J.~W. Liu, D.~Bryndin, E.~S. Božin, J.~Bloch,
  T.~Proffen, S.~J.~L. Billinge, {P}{D}{F}fit2 and {P}{D}{F}gui: computer programs for
  studying nanostructure in crystals, J. Phys.: Condens. Matter 19 (2007)
  335219.

\bibitem{gereben-2007}
O.~Gereben, P.~J{\'o}v{\'a}ri, L.~Temleitner, L.~Pusztai, {A new version of the
  RMC++ Reverse Monte Carlo programme, aimed at investigating the structure of
  covalent glasses}, Journal of Optoelectronics and Advanced Materials 9 (2007)
  3021--3027.

\bibitem{gereben-2012}
O.~Gereben, L.~Pusztai, Rmc\_pot, a computer code for reverse monte carlo
  modeling the structure of disordered systems containing molecules of
  arbitrary complexity, J. of Comp. Chem. 33 (2012) 2285--2291.

\bibitem{dove-2002}
M.~T. Dove, M.~G. Tucker, D.~A. Keen, Neutron total scattering method:
  simultaneous determination of long-range and short-range order in disordered
  materials, EUR J. MINERAL 14 (2002) 331--348.

\bibitem{Lang_1993_9_1846}
P.~L. Llewellyn, J.~P. Coulomb, Y.~Grillet, J.~Patarin, H.~Lauter, H.~Reichert,
  J.~Rouquerol, Adsorption by mfi-type zeolites examined by isothermal
  microcalorimetry and neutron diffraction. 1. argon, krypton, and methane,
  Langmuir 9 (1993) 1846 – 1851.

\bibitem{ChemRev_2008_108_4125}
B.~Smit, T.~L.~M. Maesen, Molecular simulations of zeolites: Adsorption,
  diffusion, and shape selectivity, Chem. Rev. 108 (2008) 4125--4184.

\bibitem{sanchez-2014}
V.~Sánchez-Gil, E.~G. Noya, E.~Lomba, Reverse monte carlo modeling in confined
  systems, J. Chem. Phys. 140 (2014) 024504.

\bibitem{rahman-1964}
A.~Rahman, Correlations in the motion of atoms in liquid argon, Phys. Rev. 136
  (1964) A405.

\bibitem{bellisent-funel}
Bellisent-Funel, C.~M.~C., D.~P., Levesque, J.~Weis, J, Structure factor and
  effective two-body potential for liquid gallium, Phys. Rev. A 39 (1989) 6310.

\bibitem{jovari-2001}
P.~Jóvári, L.~Pusztai, Structure of disordered forms of selenium close to the
  melting point, Phys. Rev. B 64 (2001) 014205.

\bibitem{gereben-1994}
O.~Gereben, L.~Pusztai, Structure of amorphous semiconductors: {R}everse
  {M}onte {C}arlo studies on a-{C}, a-{S}i and a-{G}e, Phys. Rev. B 50 (1994)
  14136.

\bibitem{yarnell-1973}
J.~L. Yarnell, M.~J. Katz, R.~G. Wenzel, S.~H. Koenig, Structure factor and
  radial distribution function for liquid argon at 85$^{o}${K}, Phys. Rev. A 7
  (1973) 2130.

\bibitem{l_se-data}
K.~Suzuki, Structural study of liquids with strong short-range correlation in
  the atomic distribution, Berichte der Bunsen-Gesellschaft 80~(8) (1976)
  689--694.

\bibitem{kugler-1993}
S.~Kugler, L.~Pusztai, L.~Rosta, R.~Bellisent, P.~Chieux, The structure of
  evaporated pure amorphous silicon: Neutron diffraction and reverse monte
  carlo investigations, Phys. Rev. B 48 (1993) 7685.

\bibitem{mcgreevy-2001}
R.~L. McGreevy, Reverse monte carlo modelling, J. Phys.: Condens. Matter 13
  (2001) R877--R913.

\bibitem{evrard-2005}
G.~Evrard, L.~Pusztai, {Reverse Monte Carlo modelling of the structure of
  disordered materials with RMC++ : a new implementation of the algorithm in
  C++ }, Journal of Physics: Condensed Matter 17 (2005) S1--S13.

\bibitem{white_1999}
J.~A. White, Lennard-jones as a model for argon and test of extended
  renormalization group calculations, J. Chem. Phys. 111 (1999) 9352.

\bibitem{scat_lengths}
F.~Sears, Neutron scattering llength and cross sections, Neutron News 3~(3)
  (1992) 26--37.

\end{thebibliography}

\end{document}